\documentclass[twocolumn,showpacs,preprintnumbers,amsmath,amssymb]{revtex4}

\usepackage{graphicx}
\usepackage{dcolumn}
\usepackage{bm}

\begin{document}
\draft
\title{Spectral singularities in a non-Hermitian Friedrichs-Fano-Anderson model}
\normalsize
\author{Stefano Longhi}
\address{Dipartimento di Fisica and Istituto di Fotonica e Nanotecnologie del CNR,
Politecnico di Milano, Piazza L. da Vinci 32,  I-20133 Milan,
Italy}


%
\bigskip
\begin{abstract}
\noindent Spectral singularities are predicted to occur in a
non-Hermitian extension of the Friedrichs-Fano-Anderson model
describing the decay of a discrete state $|a \rangle$ coupled to a
continuum of modes. A physical realization of the model, based on
electronic or photonic transport in a semi-infinite tight-binding
lattice with an imaginary impurity site at the lattice boundary, is
proposed. The occurrence of the spectral singularities is shown to
correspond either to a diverging reflection probability  (for an
amplifying impurity) or to a vanishing reflection probability (for
an absorbing impurity) from the lattice boundary. In the former
case, the spectral singularity of the resolvent is also responsible
for the non-decay of state $|a \rangle$ into the continuum, in spite
of the absence of bound states.
\end{abstract}



 \maketitle

\newpage

\section{Introduction.} Over the last decade, a great attention has
been devoted to investigate the properties of non-Hermitian physical
systems. In particular, it has been shown that the framework of
quantum mechanics can be extended by relaxing the common constraint
of Hermiticity for the underlying Hamiltonian $H$ (see, for
instance, \cite{Bender08,Tateo07,MostafazadehQM} and references
therein), provided that $H$ has a real energy spectrum and is
diagonalizable. In this case, after a proper change of the inner
product (metric) of the Hilbert space, the non-Hermitian Hamiltonian
$H$ may be used to define a unitary quantum system
\cite{MostafazadehQM}. Examples of such systems include
non-Hermitian Hamiltonians with parity-time ($\mathcal{PT}$)
symmetry \cite{Bender98}, in which the reality of the energy
spectrum (bound as well as radiation states) below a
symmetry-breaking transition has been proved for several complex
potentials \cite{Bender08}. Unfortunately, non-Hermitian
Hamiltonians possessing a real-valued energy spectrum may fail to be
diagonalizable because of the occurrence of exceptional points in
the point spectrum \cite{Kato,Berry}, or of spectral singularities
in the continuous part of the energy spectrum
\cite{spectral1,Samsonov05,Mostafazadeh09JPA}. Exceptional points
refer to the coalescence of two or more discrete eigenvalues
together with their eigenvectors; their physical relevance has been
investigated in several works, and different physical realizations
have been proposed and experimentally demonstrated
\cite{exceptional2}. On the other hand, spectral singularities refer
to divergences of the resolvent operator $G(z)=(z-H)^{-1}$ belonging
to the continuous spectrum of $H$, i.e. which do not correspond to
square-integrable eigenfunctions. As opposed to exceptional points,
spectral singularities have received less attention from physicists
and have been mainly viewed as a curious mathematical property of
certain non-Hermitian operators
\cite{spectral1,Samsonov05,Mostafazadeh09JPA}. The physical meaning
and relevance of spectral singularities have been highlighted solely
quite recently, notably by A. Mostafazadeh\cite{Mostafazadeh09PRL}
(see also
\cite{Samsonov05,Mostafazadeh09JPA,Mostafazadeh06JPA,Ahmed}) in the
framework of wave scattering by complex potentials \cite{Muga}.
Mostafazadeh showed that spectral singularities of a non-Hermitian
Hamiltonian with a complex potential correspond to divergences of
reflection and transmission coefficients of scattered states, i.e.
to resonances with vanishing spectral width. He also investigated in
details the appearance of spectral singularities in an
electromagnetic realization of a non-Hermitian Hamiltonian with
$\mathcal{PT}$ symmetry based on a waveguide filled by an atomic gas
\cite{Mostafazadeh09PRL,Mostafazadeh09un}, following an earlier
proposal by  Ruschhaupt and coworkers \cite{Muga2}.\\
It is the aim of this work to investigate the onset of spectral
singularities in a non-Hermitian extension of the famous
Friedrichs-Fano-Anderson (FFA) model
\cite{Friedrichs,Fano,Anderson}, which generally describes the decay
of a discrete state coupled to a continuum of modes. The FFA model
is encountered in different areas of physics, ranging from atomic
physics \cite{Piraux90,Knight90,Tannoudji} to quantum
electrodynamics \cite{Kofman94,Lambropoulos00} and condensed-matter
physics \cite{Mahan,Cini,Gadzuk00,Stefanou,Tanaka06}. The FFA model
has been studied in the quantum theory of non-integrable systems
\cite{Prigogine91} and used to describe unstable quantum systems,
quantum mechanical decay and quantum Zeno dynamics
\cite{Facchi1,Facchi2}. Simple models of single-particle electronic
or photonic transport in tight-binding lattices can be also
described by means of FFA Hamiltonians
\cite{Mahan,Cini,Gadzuk00,Stefanou,Tanaka06,Fan98,Xu00,Longhi06,Longhi07,LonghiZeno,Kivsharun}.
In this work it is shown  in particular that spectral singularities
can be observed in photonic or electronic transport in semi-infinite
tight-binding lattices with an imaginary impurity site at the
lattice boundary, leading to either a diverging or
a vanishing wave reflection from the lattice boundary.\\
 The paper
is organized as follows. In Sec.II we introduce a non-Hermitian
extension of the FFA model, and derive the conditions for the
appearance of spectral singularities. In particular, it is shown by
direct calculations that the appearance of divergences of the
resolvent operator $G(z)$ on the branch cut correspond to the
non-diagonalizability of the FFA Hamiltonian. In Sec.III we present
an example of non-Hermitian FFA model showing spectral
singularities, which provides a simple model of electron or photonic
transport in a semi-infinite tight-binding lattice with a boundary
impurity site. It is shown that spectral singularities in this model
correspond to either a vanishing or a diverging  reflection
probability from the lattice boundary. Finally, in Sec.IV the main
conclusions are outlined.

\section{Spectral singularities in a non-Hermitian Friedrichs-Fano-Anderson model}

\subsection{The model}
The standard FFA model describes the interaction of a discrete state
$|a\rangle$, of energy $E_a$, with a continuous set of states
$|k\rangle$ with energy $ E(k)$ (see, for instance,
\cite{Tannoudji,Lambropoulos00,Mahan,Cini,Tanaka06,Longhi07}). Here
$|a\rangle$ and $|k\rangle$ represent a complete set of Dirac states
in the Hilbert space which satisfy the orthonormal conditions
$\langle a | a \rangle=1$, $\langle a | k \rangle=0$ and $\langle k'
| k \rangle= \delta(k-k')$. The Hamiltonian of the full system can
be written as $H=H_0+V$, where
\begin{equation}
H_0=E_a |a\rangle\langle a |+ \int dk \; E(k) |k \rangle\langle k |
\end{equation}
is the Hamiltonian of the non-interacting discrete and continuous
states, and
\begin{equation}
V=\int dk  \left[ v(k) |a \rangle\langle k |+v^*(k) |k
\rangle\langle a | \right]
\end{equation}
is the (self-adjoint) interaction term, described by the spectral
coupling function $v(k)$. Typically, we assume that the energy $E
(k)$ of continuous states spans the interval $E_1 < E < E_2$
(eventually $E_2=\infty$), and $E(k)$ is a monotonic function
(either increasing or decreasing) of $k$, i.e. we assume that there
are not energy degeneracies of continuous states. The state vector
of the system $|\psi\rangle$ evolves according to the
Schr\"{o}dinger equation (with $\hbar=1$)
\begin{equation}
i \frac{\partial |\psi\rangle}{\partial t}=H |\psi\rangle.
\end{equation}
Note that, provided that the energies $E_a$ and $E(k)$ are
real-valued, the Hamiltonian $H$ is Hermitian. The spectrum of $H$ can
be determined by either analyzing the singularities of the resolvent
$G(z)=(z-H)^{-1}$, or by projecting the eigenvalue equation $H
|\psi\rangle=E|\psi\rangle$ on the basis $\{ |a\rangle ,
|k\rangle\}$. As is well known, the continuous spectrum of $H$ is
 $E_1 < E <  E_2$, i.e. the same as that of $H_0$, whereas the
 point spectrum can be either empty or composed by a number of discrete
 eigenvalues, either outside or embedded into the continuous
 spectrum (see, for instance, \cite{Tannoudji,Sudarshan,Miyamoto05}).\\
We now relax the Hermiticity condition of the FFA model by allowing
the 'energy' $E_a$ of the discrete state $|a\rangle$ to be complex
valued. However, we will assume that the spectrum of $H$ remains
real-valued in spite of the non-Hermiticity of $H$. The condition
for the spectrum of $H$ to remain real-valued will be discussed
below in Sec.II.B and corresponds to the absence of bound states,
i.e. to an empty point spectrum. This means that a non-Hermitian FFA
Hamiltonian has a real-valued energy spectrum if and only if its
spectrum is purely continuous. We are interested here to determine,
if any, the appearance of spectral singularities of $H$, which would
prevent $H$ to be diagonalizable. This problem can be addressed in
two ways: (i) by the determination of the resolvent $G(z)$, and (ii)
by a direct diagonalization of $H$ following the original procedure
by Fano \cite{Fano}, extended to account for the non-Hermitian
nature of $H$. In the former case, a spectral singularity at
$\mathcal{E}=\mathcal{E}_0$, embedded in the continuous spectrum
$(E_1,E_2)$, is revealed as a divergence of the Green function
$\mathcal{G}(x,y;z)=\langle x | G(z) y \rangle$ in a neighborhood of
$z=\mathcal{E}_0$, divergence which does not corresponds to a bound
state embedded in the continuum \cite{Mostafazadeh09JPA}. In the
latter case, a spectral singularity at $\mathcal{E}=\mathcal{E}_0$
occurs when $\langle \mathcal{E}_0^{\dag}|\mathcal{E}_0\rangle=0$,
where $|\mathcal{E}_0 \rangle$ and $|\mathcal{E}^{\dag}_0 \rangle$
are the eigenfunctions of $H$ and of its adjoint $H^{\dag}$,
respectively, corresponding to the eigenvalue $\mathcal{E}_0$
\cite{Mostafazadeh09JPA}.

\subsection{Spectral singularities: the resolvent approach}
For a given complex number $z$, the resolvent operator $G(z)$ of the
Hamiltonian $H$ is defined as
\begin{equation}
G(z)=(z-H)^{-1},
\end{equation}
i.e. $G(z)(z-H)=(z-H)G(z)=\mathcal{I}$, where $\mathcal{I}$ is the
identity operator. The knowledge of the resolvent of $H$, for any
$z$, is equivalent to the knowledge of the set of eigenfunctions and
eigenvalues of $H$. In particular, the singularities of $G(z)$ in
the complex plane define the spectrum of $H$: an eigenvalue
$\mathcal{E}$ belonging to the point spectrum of $H$ is a pole of
$G(z)$, whereas the branch cut of $G(z)$ determines the continuous
part of the spectrum of $H$. The Hamiltonian $H$ is said to have a
spectral singularity at $\mathcal{E}=\mathcal{E}_0$, where
$\mathcal{E}_0$ belongs to the continuous spectrum of $H$, if the
function
\begin{equation}
\mathcal{G}_{\chi,\varphi}(z) =\langle \chi |G(z) \varphi \rangle
\end{equation}
 is unbounded in the neighborhood of
$z=\mathcal{E}_0$, and $\mathcal{E}_0$ does not belong to the point
spectrum of $H$, i.e. it does not corresponds to a bound state
embedded in the continuum \cite{Mostafazadeh09JPA}. In the previous
equation, $|\chi\rangle$ and $|\varphi\rangle$ are two assigned
functions of the Hilbert space; in particular, for $|\chi\rangle=|x
\rangle$ and $|\varphi\rangle=|y \rangle$ one obtains the coordinate
representation of the resolvent $G$, i.e. the Green function
$\mathcal{G}(x,y;z)=\langle x |G(z) y \rangle$. An interesting
property of the FFA Hamiltonian is the possibility to calculate the
resolvent in a closed form. The procedure to calculate $G(z)$ is
well known for the Hermitian case (see, for instance, \cite{Cini}),
and can be extended {\it mutatis mutandis} to the non-Hermitian FFA
model considered in this work. As detailed in the Appendix A, the
matrix elements of the resolvent $G(z)$ on the complete basis $\{
|a\rangle, |k\rangle \}$ read explicitly
\begin{widetext}
\begin{eqnarray}
\mathcal{G}_{a,a}  \equiv  \langle a | G(z) a \rangle & = &
\frac{1}{z-E_a-\Sigma(z)} \\
\mathcal{G}_{a,k}  \equiv  \langle a | G(z) k \rangle & = &
\frac{v(k)}{(z-E(k))(z-E_a-\Sigma(z))} \\
\mathcal{G}_{k,a}  \equiv  \langle k | G(z) a \rangle & = &
\frac{v^*(k)}{(z-E(k))(z-E_a-\Sigma(z))} \\
\mathcal{G}_{k,k'}  \equiv  \langle k | G(z) k' \rangle & = &
\frac{v(k')
v^*(k)}{(z-E(k))(z-E(k'))(z-E_a-\Sigma(z))}+\frac{\delta(k-k')}{z-E(k')}
\end{eqnarray}
\end{widetext}
where $\Sigma(z)$ is the self-energy, defined by
\begin{equation}
\Sigma(z)= \int dk \frac{|v(k)|^2}{z-E(k)}.
\end{equation}
Note that, by introducing the density of states $\rho(E)=(\partial E
/ \partial k)^{-1}$ and letting $V(E)=\rho(E)|v(E)|^2$ ($V(E)$=0 for
$E>E_2$ and $E<E_1$), the self-energy can be written in the
equivalent form
\begin{equation}
\Sigma(z)= \int_{E_1}^{E_2} dE \frac{V(E)}{z-E}.
\end{equation}
Note also that $\Sigma(z)$ is not defined on the segment $(E_1,E_2)$
of the real axis, and one has
\begin{equation}
\Sigma(z=\mathcal{E} \pm i 0^+)= \Delta(\mathcal{E}) \mp i \pi
V(\mathcal{E})
\end{equation}
where
\begin{equation}
\Delta(\mathcal{E})=P \int_{E_1}^{E_2} dE
\frac{V(E)}{\mathcal{E}-E},
\end{equation}
$P$ denotes the principal value, and $\mathcal{E}$ is real-valued.
Therefore, $\Sigma(z)$ has a branch cut in the interval $(E_1,E_2)$.

For two assigned functions $|\chi \rangle=\chi_a |a\rangle + \int dk
\chi(k) |k\rangle $ and $|\varphi \rangle=\varphi_a |a\rangle + \int
dk \varphi(k) |k\rangle$ of the Hilbert space, the complex function
$\mathcal{G}_{\chi, \varphi}(z)=\langle \chi | G(z) \varphi \rangle$
can be readily calculated as
\begin{widetext}
\begin{equation}
\mathcal{G}_{\chi, \varphi}(z)=\chi_{a}^* \varphi_a
\mathcal{G}_{a,a}(z)+\chi_{a}^* \int dk \varphi(k)
\mathcal{G}_{a,k}(z)+ \varphi_a \int dk \chi^*(k)
\mathcal{G}_{k,a}(z) + \int dk dk' \chi^{*}(k') \varphi(k)
\mathcal{G}_{k',k}(z)
\end{equation}
\end{widetext}
Substitution of Eqs.(6-9) into Eq.(14) finally yields
\begin{equation}
\mathcal{G}_{\chi, \varphi}(z) =  \mathcal{G}_{a,a}(z) \Phi_1(z)+
\Phi_2(z)
\end{equation}
where we have set
\begin{widetext}
\begin{eqnarray}
\Phi_1(z) & = & \chi^{*}_a \varphi_a+\chi^{*}_a \int dE
\frac{\rho(E) \varphi(E) v(E)}{z-E} +\varphi_a \int dE \frac{\rho(E)
\chi^*(E) v^*(E)}{z-E}+ \nonumber \\
& + & \left( \int dE \frac{\rho(E) v(E)\varphi(E)}{z-E}\right)
\left( \int dE
\frac{\rho(E) v^*(E)\chi^*(E)}{z-E}\right) \\
\Phi_2(z) & = & \int dE \frac{\rho(E) \chi^*(E) \varphi(E)}{z-E}.
\end{eqnarray}
\end{widetext}
We are now ready to determine the singularities of the resolvent,
i.e. the spectrum of $H$ and possible spectral singularities. To
this aim, let us notice that $\Phi_1(z)$ and $\Phi_2(z)$ are bounded
functions of $z$ and have a branch cut on the segment $(E_1,E_2)$ of
the real axis (as for the self-energy $\Sigma$). According to
Eq.(15), we may therefore limit to consider the singularities of
$\mathcal{G}_{a,a}(z)$. From
Eqs.(6), (11) and (12) one can conclude that:\\
(i) The continuous spectrum of $H$ is the same as that of $H_0$,
i.e. the interval $(E_1,E_2)$ of the real axis, where $G(z)$ has a branch cut. \\
(ii) The point spectrum of $H$, corresponding to bound states
outside the continuum, are the complex roots $z$ of the equation
\begin{equation}
z-E_a=\Sigma(z)
\end{equation}
in correspondence of which the resolvent $G(z)$ has a pole.\\
(iii) A spectral singularity $\mathcal{E}_0$ of the continuous
spectrum$(E_1,E_2)$ is any solution of the coupled equations
\begin{eqnarray}
{\rm Im}(E_a)  & = &  \pm \pi V(\mathcal{E}_0) \\
\mathcal{E}_0-{\rm Re}(E_a) & = &\Delta(\mathcal{E}_0)
\end{eqnarray}
provided that ${\rm Im}(E_a) \neq 0$ \cite{note1}.\\
Note that real-valued energies can not belong to the point spectrum
of $H$ because Eq.(18) does not have real-valued roots whenever $H$
is non-Hermitian \cite{note2}. This means that the point spectrum of
$H$, if not empty, is strictly complex-valued. On the other hand,
the continuous part of the spectrum is real-valued according to the
property (i). Therefore, we may conclude that the energy spectrum of
the non-Hermitian FFA Hamiltonian is real-valued if and only if its
spectrum is purely continuous. In this case, spectral singularities
in the continuous spectrum occur whenever Eqs.(19) and (20) can be
simultaneously satisfied. It should be noted that the behavior of
$G(z)$ in the neighborhood of the spectral singularity
$z=\mathcal{E}_0$ on the continuous spectrum is different for an
'absorbing' [$\mathrm{Im}(E_a)<0$] and for an 'amplifying'
[$\mathrm{Im}(E_a)>0$] complex energy $E_a$ of state $|a\rangle$.
Since the spectral coupling $V(\mathcal{E}_0)$ is always positive,
for an absorbing complex energy Eq.(19) can be satisfied by taking
the lower (negative) sign on the right hand side; correspondingly,
from Eqs.(6) and (12) it follows that $\mathcal{G}_{a,a}(z)$ is
unbounded when $z \rightarrow \mathcal{E}_0$ from the bottom of the
real energy axis, i.e. for $\mathrm{Im}(z)<0$, {\it but }
$\mathcal{G}_{a,a}(z)$ remains bounded when $z \rightarrow
\mathcal{E}_0$ with $\mathrm{Im}(z)>0$ \cite{noteIMP}. Conversely,
for an 'amplifying' complex energy $\mathrm{Im}(E_a)>0$, Eq.(19) can
be satisfied by taking the upper (positive) sign on the right hand
side; correspondingly, $\mathcal{G}_{a,a}(z)$ is unbounded when $z
\rightarrow \mathcal{E}_0$  with $\mathrm{Im}(z)>0$, {\it but} it
remains bounded when $z \rightarrow \mathcal{E}_0$ with
$\mathrm{Im}(z)<0$ \cite{noteIMP}. Such a different behavior of
spectral singularities for an absorbing or an amplifying complex
energy $E_a$ has some relevant physical implications, that will be
discussed in Sec.II.D and Sec.III.C. In Sec.II.D it will be shown
that the appearance of a spectral singularity in the amplifying case
is responsible for the non-decay of state $|a \rangle$ into the
continuum, in spite of the absence of bound states; in Sec.III.C the
interplay between spectral singularities and wave scattering will be
investigated for a semi-infinite tight-binding lattice realization
of the FFA Hamiltonian.

\subsection{Spectral singularities: the Fano diagonalization procedure}
Let us assume that the spectrum of the non-Hermitian Hamiltonian
$H=H_0+V$ is real-valued. As shown in the previous subsection, this
implies that the spectrum of $H$ is purely continuous and spans the
interval $(E_1,E_2)$. Let us indicate by $|\mathcal{E}\rangle$ the
(improper) eigenfunction of $H$ corresponding to the eigenvalue
$\mathcal{E}$, and by $|\mathcal{E}^{\dag}\rangle$ the (improper)
eigenfunction of the adjoint $H^{\dag}$ corresponding to the same
eigenvalue $\mathcal{E}$. For $H$ to be diagonalizable, the set of
functions $\{ | \mathcal{E} \rangle, |\mathcal{E}^{\dag} \rangle \}$
must form a complete biorthonormal basis of Hilbert space
\cite{Mostafazadeh09JPA}, that is
\begin{equation}
\langle \mathcal{E}^{\dag}|\mathcal{E}' \rangle =
\delta(\mathcal{E}-\mathcal{E}') \;, \; \; \int_{E_1}^{E_2}
d\mathcal{E} |\mathcal{E} \rangle \langle
\mathcal{E}^{\dag}|=\mathcal{I}.
\end{equation}
A spectral singularity at $\mathcal{E}=\mathcal{E}_0$ sets in when
\begin{equation}
\langle \mathcal{E}_0|\mathcal{E}_{0}^{\dag} \rangle=0,
\end{equation}
 which prevents $H$ to be diagonalizable. To determine the onset of spectral
 singularities of the FFA Hamiltonian $H$, we can thus proceed by
 calculating the eigenfunctions $|\mathcal{E} \rangle$ of $H$ and $|\mathcal{E}^{\dag} \rangle$ of its adjoint $H^{\dag}$
 following the diagonalization procedure used by Fano
 \cite{Fano} in the problem of atomic autoionizing resonances (see also
 \cite{Friedrichs,Knight90}), properly modified to account for the non-Hermitian nature of $H$. To this aim, we expand the eigenstates
 $|\mathcal{E} \rangle$ and $|\mathcal{E}^{\dag} \rangle$ on the
 orthonormal and complete basis $\{ |a \rangle, |k
 \rangle \}$ as
 \begin{eqnarray}
|\mathcal{E} \rangle & = &  \alpha(\mathcal{E}) |a\rangle +
\int dk \beta(\mathcal{E},k) |k \rangle \\
|\mathcal{E}^{\dag} \rangle & = &  \alpha^{\dag}(\mathcal{E})
|a\rangle + \int dk \beta^{\dag}(\mathcal{E},k) |k \rangle
 \end{eqnarray}
with expansion coefficients $\alpha(\mathcal{E})$,
$\alpha^{\dag}(\mathcal{E})$, $\beta(\mathcal{E},k)$ and
$\beta^{\dag}(\mathcal{E},k)$ to be determined. Note that
\begin{equation}
\langle \mathcal{E}'|\mathcal{E}^{\dag}
\rangle=\alpha^*(\mathcal{E}')\alpha^{\dag}(\mathcal{E})+ \int dk
\beta^{*}(\mathcal{E}',k) \beta^{\dag}(\mathcal{E},k).
\end{equation}
Using Eqs.(1), (2) and (23), the eigenvalue equation $H|\mathcal{E}
\rangle=\mathcal{E}|\mathcal{E} \rangle$ yields the following
coupled equations for the expansion coefficients
$\alpha(\mathcal{E})$ and $\beta(\mathcal{E},k)$:
\begin{eqnarray}
(E_a-\mathcal{E}) \alpha(\mathcal{E})+\int dk v(k)
\beta(\mathcal{E},k)=0 \\
\left[ E(k)-\mathcal{E} \right] \beta(\mathcal{E},k)+v^*(k)
\alpha(\mathcal{E})=0.
\end{eqnarray}
Equation (27) can be solved for $\beta(\mathcal{E},k)$ and
substituted into Eq.(26). However, there is a singularity at
$E(k)=\mathcal{E}$, so that $1/ \left[E(k)-\mathcal{E} \right]$ must
be written as its principal and delta function parts, yielding (see
\cite{Fano,Knight90})
\begin{equation}
\beta(\mathcal{E},k)=-v^{*}(k) \alpha(\mathcal{E}) \left( {\rm P}
\frac{1}{E(k)-\mathcal{E}} +z(\mathcal{E}) \delta(E(k)-\mathcal{E})
\right).
\end{equation}
The coefficient $z(\mathcal{E})$ in front of the delta function on
the right hand side of Eq.(28) is determined by substituting Eq.(28)
into Eq.(26) and requiring that $\alpha(\mathcal{E})$ does not
vanish. This yields
\begin{equation}
z(\mathcal{E})=\frac{E_a-\mathcal{E}+\Delta(\mathcal{E})}{V(\mathcal{E})}
\end{equation}
where $\Delta(\mathcal{E})$ is defined by Eq.(13),
$V(\mathcal{E})=\rho(\mathcal{E})|v(\mathcal{E})|^2$ and
$\rho(\mathcal{E})=(\partial E(k) / \partial k)^{-1}$ is the density
of states.\\
 Similarly, the eigenvalue equation
$H^{\dag}|\mathcal{E}^{\dag} \rangle=\mathcal{E}|\mathcal{E}^{\dag}
\rangle$ yields
\begin{eqnarray}
(E_{a}^{*}-\mathcal{E}) \alpha^{\dag}(\mathcal{E})+\int dk v(k)
\beta^{\dag}(\mathcal{E},k)=0 \\
\left[ E(k)-\mathcal{E} \right] \beta^{\dag}(\mathcal{E},k)+v^*(k)
\alpha^{\dag}(\mathcal{E})=0
\end{eqnarray}
and the following expression of the coefficient
$\beta^{\dag}(\mathcal{E},k)$ can be derived following the same
procedure as above
\begin{equation}
\beta^{\dag}(\mathcal{E},k)=-v^{*}(k) \alpha^{\dag}(\mathcal{E})
\left( {\rm P} \frac{1}{E(k)-\mathcal{E}} +z^{*}(\mathcal{E})
\delta(E(k)-\mathcal{E}) \right).
\end{equation}
Substitution of Eqs.(28) and (32) into Eq.(25) yields
\begin{equation}
\langle \mathcal{E}'|\mathcal{E}^{\dag} \rangle=
\alpha^*(\mathcal{E}') \alpha^{\dag}(\mathcal{E})
F(\mathcal{E},\mathcal{E}')
\end{equation}
where we have set
\begin{eqnarray}
F(\mathcal{E},\mathcal{E}') & = & \int_{E_1}^{E_2}dE V(E) \left(
{\rm P} \frac{1}{E-\mathcal{E}} +z^*(\mathcal{E})
\delta(E-\mathcal{E})\right) \nonumber \\
& \times &  \left( {\rm P} \frac{1}{E-\mathcal{E}'}
+z^*(\mathcal{E}') \delta(E-\mathcal{E}')\right)+1.
\end{eqnarray}
The calculation of the integral on the right hand side of Eq.(34) is
complicated by the product of two principal parts, which must be
properly resolved into its partial-fraction and singular terms (see
\cite{Fano} or the Appendix of Ref.\cite{Knight90}). Taking into
account that
\begin{eqnarray}
{\rm P} \frac{1}{E-\mathcal{E}} {\rm P} \frac{1}{E-\mathcal{E}'} & =
& {\rm P} \frac{1}{\mathcal{E}-\mathcal{E}'} \left( {\rm P}
\frac{1}{\mathcal{E}'-E}-{\rm P} \frac{1}{\mathcal{E}-E} \right)+
\nonumber \\
& + & \pi^2 \delta(\mathcal{E}'-E) \delta(\mathcal{E}-E)
\end{eqnarray}
and using Eq.(29), from Eq.(34) one obtains
\begin{equation}
F(\mathcal{E},\mathcal{E}')=V(\mathcal{E}) \left[
\pi^2+z^{*2}(\mathcal{E})\right] \delta(\mathcal{E}-\mathcal{E}')
\end{equation}
so that [see Eq.(33)]
\begin{equation}
\langle \mathcal{E}'|\mathcal{E}^{\dag} \rangle=
\alpha^*(\mathcal{E}) \alpha^{\dag}(\mathcal{E}) V(\mathcal{E})
\left[ \pi^2+z^{*2}(\mathcal{E})\right]
\delta(\mathcal{E}-\mathcal{E}').
\end{equation}
In order $\{ |\mathcal{E} \rangle,  |\mathcal{E}^{\dag} \rangle \}$
to represent a complete biorthonormal set of functions [Eq.(21)],
the amplitudes $\alpha(\mathcal{E})$ and
$\alpha^{\dag}(\mathcal{E})$ should be thus normalized such that
\begin{equation}
\alpha^*(\mathcal{E}) \alpha^{\dag}(\mathcal{E}) V(\mathcal{E})
\left[ \pi^2+z^{*2}(\mathcal{E})\right]=1.
\end{equation}
For the Hermitian FFA model, the energy $E_a$ is real-valued,
$z(\mathcal{E})$ turns out to be real-valued and
$\alpha^{\dag}(\mathcal{E})=\alpha(\mathcal{E})$ [because of
$|\mathcal{E}^{\dag} \rangle =|\mathcal{E} \rangle$], so that
Eq.(38) is solved to yield
\begin{equation}
|\alpha(\mathcal{E})|^2=\frac{V(\mathcal{E})}{\pi^2
V^2(\mathcal{E})+[E_a-\mathcal{E}+\Delta(\mathcal{E})]^2}
\end{equation}
according to Fano \cite{Fano}. The physical meaning of Eq.(39) is
well known \cite{Fano}: owing to the coupling with the continuum,
the discrete state $|a\rangle$ is 'diluited' throughout a set of
continuous states (i.e., it becomes a resonance for $H$) with a
typical resonance curve $|\alpha(\mathcal{E})|^2$ peaked at
$\mathcal{E} \simeq E_a-\Delta(E_a)$ of width $ \simeq \pi V(E_a)$
\cite{note3}.\\
For the non-Hermitian FFA model, the energy $E_a$ is complex-valued
and from Eq.(37) it follows that a spectral singularity can appear
at the energy $\mathcal{E}=\mathcal{E}_0$ of the continuous spectrum
such that $\pi^2+z^{*2}(\mathcal{E}_0)=0$, i.e. when
\begin{equation}
z^{*}(\mathcal{E}_0)= \mp i \pi.
\end{equation}
Using Eq.(29), Eq.(40) yields the following conditions for the
appearance of a spectral singularity in the non-Hermitian FFA model
\begin{equation}
\mathrm{Im}(E_a) = \pm \pi V(\mathcal{E}_0) \; , \;
\mathrm{Re}(E_a)-\mathcal{E}_0+\Delta(\mathcal{E}_0)=0
\end{equation}
which are precisely Eqs.(19) and (20) derived in Sec.II.B following
the resolvent approach to spectral singularities.

\subsection{Spectral singularities and decay dynamics}
Hermitian FFA models are often used to describe the evolution of
unstable quantum systems and the related problem of quantum
mechanical decay and deviations from an exponential decay law (see,
for instance, \cite{Facchi1,Facchi2} and references therein). If the
system is initially prepared in state $|a \rangle$, i.e. if
$|\psi(t=0) \rangle=|a \rangle$, the survival probability $P(t)$ to
find the system at state $|a \rangle$ at a successive time $t$ is
given by $P(t)=|c_a(t)|^2$, where
\begin{equation}
c_a(t)= \langle a | \psi(t) \rangle = \langle a | \exp(-iHt) |a
\rangle.
\end{equation}
Here we consider the case of a non-Hermitian FFA Hamiltonian $H$
with a real-valued (and thus purely continuous) energy spectrum, and
briefly discuss the physical implications of spectral singularities
of $H$ on the decay dynamics of survival probability $P(t)$. The
temporal evolution operator $ \exp(-iHt)$ of the system can be
written in terms of the resolvent $G(z)$ as
\begin{equation}
\exp(-iHt)=\frac{i}{2 \pi} \int_{\rm B} dz G(z) \exp(-izt)
\end{equation}
where the Bromwich path B is any horizontal line $\mathrm{Im}(z)=
\mathrm{const}>0$ in the upper half of the complex $z$ plane [see
Fig.1(a)]. Substitution of Eq.(43) into Eq.(42) yields
\begin{equation}
c_a(t)=\frac{i}{2 \pi} \int_{\rm B} dz \mathcal{G}_{a,a}(z)
\exp(-izt)
\end{equation}
where the matrix element $\mathcal{G}_{a,a}(z)$ of the resolvent is
given by Eq.(6). As discussed in Sec.II.B, $\mathcal{G}_{a,a}(z)$ is
analytic in the full complex plane, expect for the branch cut on the
real axis, from $z=E_1$ to $z=E_2$ (see Fig.1), where it can also
become unbounded at energies $\mathcal{E}_0$ corresponding to
spectral singularities. The integral on the right hand side of
Eq.(44) can be evaluated by the residue method after suitably
closing the Bromwich path B with a contour in the $\mathrm{Im}(z)<0$
half-plane (see, e.g., \cite{Facchi1}). Since the closure crosses
the branch cut on the real axis, the contour must necessarily pass
into the second Riemannian sheet in the sector of the
$\mathrm{Im}(z)<0$ half-plane with $E_1 < \mathrm{Re}(z)<E_2$,
whereas it remains in the first Riemannian sheet in the other two
sectors $\mathrm{Re}(z)<E_1$ and $\mathrm{Re}(z)>E_2$ of the
$\mathrm{Im}(z)<0$ half-plane. To properly close the contour, it is
thus necessary to go back and turn around the two branch points of
the cut at $z=E_1$  and $z=E_2$, following the Hankel paths $h_1$
and $h_2$ as shown in Fig.1(b). Indicating by
$\mathcal{G}_{a,a}^{(II)}(z)$ the analytic continuation of
$\mathcal{G}_{a,a}(z)$ from the upper [$\mathrm{Im}(z)>0$] to the
lower [$\mathrm{Im}(z)<0$] half-plane across the branch cut, and by
$z_1$, $z_2$, ... the (possible) poles of
$\mathcal{G}_{a,a}^{(II)}(z)$ that lies in the sector
$\mathrm{Im}(z)< \eta$, $ E_1<\mathrm{Re}(z)< E_2$ of the complex
plane (with $\eta>0$ arbitrarily small), assuming that the poles are
of first order one can write
\begin{equation}
c_a(t)=\sum_{z_k} R_k \exp(-iz_k t)+\mathcal{C}(t)
\end{equation}
\begin{figure}
\includegraphics[scale=0.43]{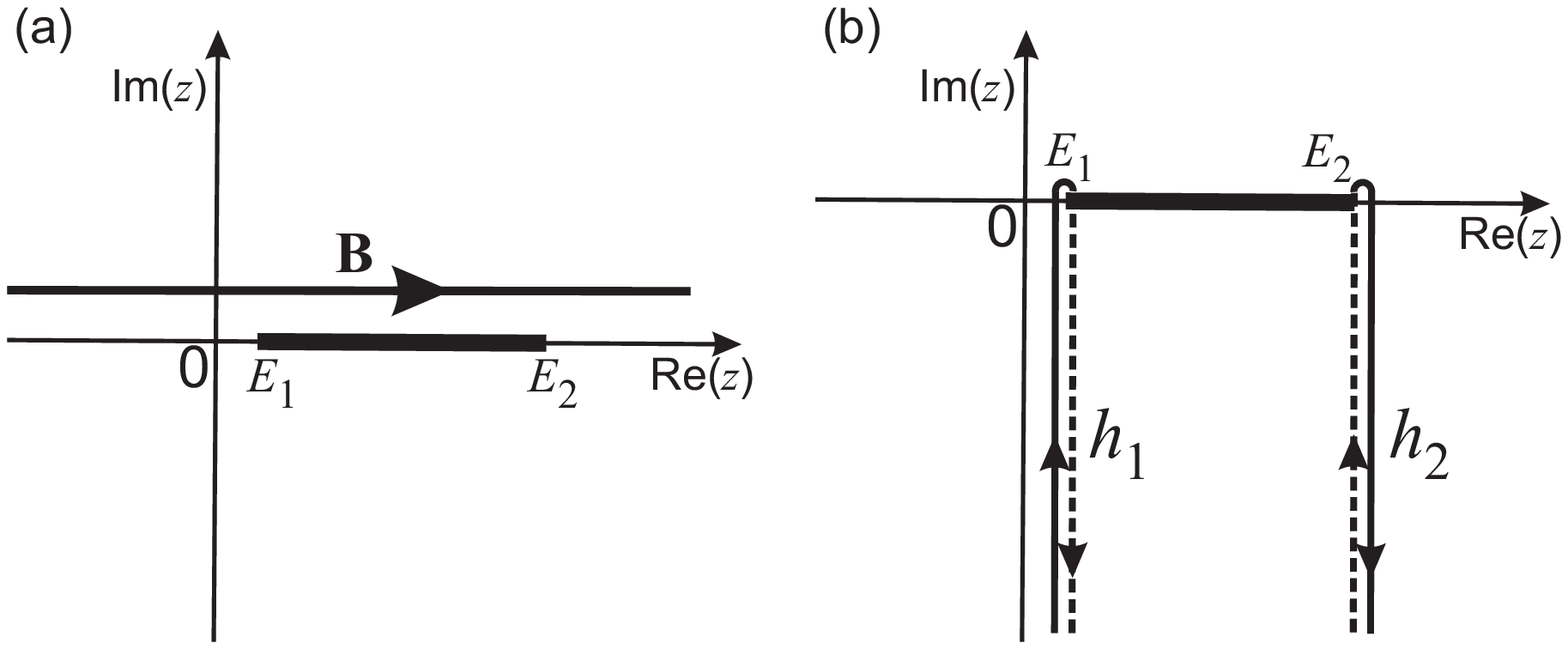} \caption{
(a) Integration contour (Bromwich path ${\mathrm B}$) in the
$\mathrm{Im}(z)>0$ complex plain entering in Eqs.(43) and (44).The
bold horizontal segment on the real axis is the continuous spectrum
of $H$ and corresponds to a branch cut of $\mathcal{G}_{a,a}(z)$.
(b) Integration contour (Hankel paths $h_1$ and $h_2$) after
deformation of the Bromwich path. The integration along the solid
(dashed) curves of the Hankel paths is made on the first (second)
Riemannian sheet of
 $\mathcal{G}_{a,a}(z)$.}
\end{figure}
where $R_k$ is the residue of $\mathcal{G}_{a,a}^{(II)}(z)$ at
$z=z_k$, and $\mathcal{C}(t)$ is the contribution from the contour
integration along the Hankel paths $h_1$ and $h_2$
\begin{eqnarray}
\mathcal{C}(t) & = &  \frac{i}{2 \pi} \int_{E_1-i \infty}^{E_1+i0}
dz [\mathcal{G}_{a,a}(z)-\mathcal{G}_{a,a}^{(II)}(z)] \exp(-izt)+
\nonumber \\
& + & \frac{i}{2 \pi} \int_{E_2-i \infty}^{E_2+i0} dz
[\mathcal{G}_{a,a}^{(II)}(z)-\mathcal{G}_{a,a}(z)] \exp(-izt).
\end{eqnarray}
The cut contribution $\mathcal{C}(t)$ vanishes as $t \rightarrow
\infty$, however it is responsible for the appearance of
nonexponential features in the decay dynamics \cite{Facchi1}. For
the Hermitian FFA model, the poles $z_k$ of
$\mathcal{G}_{a,a}^{(II)}(z)$ lie below the real axis, i.e.
$\mathrm{Im}(z_k)<0$, because $\mathcal{G}_{a,a}(z=E+i0^+)$ is a
bounded function. Therefore $P(t) \rightarrow 0$ as $t \rightarrow
\infty$ \cite{noteq}. Typically, $\mathcal{G}_{a,a}^{(II)}(z)$ has
one pole below the imaginary axis \cite{Facchi1}, so that the decay
of $c_a(t)$ follows an exponential law (with decay rate determined
by the imaginary part of the pole), corrected by the cut
contribution $\mathcal{C}(t)$. Let us consider now the non-Hermitian
FFA model with a spectral singularity at $\mathcal{E}=\mathcal{E}_0$
in the continuous spectrum. In this case, we have to distinguish two
cases. For an absorbing complex energy $E_a$, as shown in Sec.II.B
$\mathcal{G}_{a,a}(z)$ is bounded for $z=E+i0^+$; therefore, its
analytic continuation $\mathcal{G}_{a,a}^{(II)}(z)$ in the
lower-half complex plane is bounded for $z=E-i0^+$. This means that
possible poles of $\mathcal{G}_{a,a}^{(II)}(z)$ have a strictly
negative imaginary part, as for an Hermitian FFA model. Therefore
the survival probability $P(t)$ decays toward zero as $t \rightarrow
\infty$ similarly to what happens in the Hermitian limit.
Conversely, for an amplifying complex energy $E_a$,
$\mathcal{G}_{a,a}(z)$ diverges for $z=\mathcal{E}_0+i0^+$;
therefore $\mathcal{G}_{a,a}^{(II)}(z)$ has a pole on the real axis,
$z_1=\mathcal{E}_0$. In this case, according to Eq.(45) the survival
probability $P(t)$ does not decay, and may become larger than one
depending on the modulus of the residue of the dominant pole. This
phenomenon is analogous to the fractional decay found in Hermitian
FFA models and related to the existence of bound states (i.e. a
non-empty point spectrum of $H$). However, in the non-Hermitian FFA
model with $\mathrm{Im}(E_a)>0$ the non-decaying behavior of $P(t)$
results from the appearance of a spectral singularity in the
continuous spectrum, not from the existence of bound states. It
should be noted that such a non-decaying behavior was previously
predicted in the study of the lasing threshold of an optical
microcavity resonantly coupled to a coupled resonator optical
waveguide under special coupling conditions \cite{Longhi06}. This
unusual behavior of laser phase transition was explained as a
non-markovian effect arising from the structured continuum of the
decay channel, however it was not related to spectral singularities
of the underlying Hamiltonian.

\section{Spectral singularities in a semi-infinte tight-binding lattice with a boundary impurity site}
In this section we present a simple and analytically-solvable
example of a non-Hermitian FFA model showing spectral singularities,
which describes rather generally single-particle electron or photon
transport on a semi-infinite one-dimensional tight-binding lattice
with an impurity site. The model is first presented in the framework
of the general theory developed in Sec.II, and its tight-binding
lattice realization is subsequently described. The physical
implications of spectral singularities on wave scattering from the
lattice boundary and on the decay dynamics of the impurity site are
finally highlighted.

\subsection{The Hamiltonian}
Let us consider the non-Hermitian FFA model defined by the following
relations for the energy dispersion $E(k)$ and spectral coupling
$v(k)$
\begin{equation}
E(k)=-2 \kappa_0 \cos k \; , \; v(k)= -\sqrt{\frac{2}{\pi}} \kappa_a
\sin k
\end{equation}
where $\kappa_0$, $\kappa_a$ are two real-valued positive constants
and $0 \leq k \leq \pi$. The Hermitian limit of this model, attained
by assuming $\mathrm{Im}(E_a)=0$, is a special case of the FFA model
previously investigated in Ref.\cite{Longhi07}, which is exactly
solvable (see also \cite{Tanaka06}). Note that the continuous
spectrum of $H$ spans the band $(E_1,E_2)$, with $E_2=-E_1=2
\kappa_0$. The density of states for this model is given by
\begin{equation}
\rho(E)= \left( \frac{\partial E}{\partial k}\right)^{-1} =\left\{
\begin{array}{c c}
 \frac{1}{\sqrt{4 \kappa_0^2-E^2}} & -2 \kappa_0 <
E < 2 \kappa_0 \\
0 & |E|>2 \kappa_0
\end{array}
\right.
\end{equation}
which shows van-Hove singularities at the band edges, whereas the
positive spectral function $V(E)$, defined by
$V(E)=\rho(E)|v(E)|^2$, reads
\begin{equation}
V(E)= \left\{
\begin{array}{cc}
\frac{\kappa_a^2}{\pi \kappa_0} \sqrt{1-\left(\frac{E}{2
\kappa_0}\right)^2} & -2 \kappa_0 <
E < 2 \kappa_0 \\
0 & |E|>2 \kappa_0
\end{array}
\right.
\end{equation}
which is non-singular. Substitution of Eq.(49)  into Eq.(11) yields
the following expression for the self-energy $\Sigma(z)$
\cite{note4}
\begin{equation}
\Sigma(z)=-i \frac{\kappa_a^2}{2\kappa_0^2}
\left(\sqrt{4\kappa_0^2-z^2}+iz \right)
\end{equation}
and thus [see Eq.(12)]
\begin{eqnarray}
\Delta(\mathcal{E}) & = & \mathrm{Re} \left(\Sigma(z=\mathcal{E} \pm
i 0^+) \right)= \\ & = & \left\{
\begin{array}{ll}
\frac{\kappa_a^2}{2 \kappa_0^2} \left(
\mathcal{E}+\sqrt{\mathcal{E}^2-4 \kappa_0^2} \right) &
\mathcal{E}<-2\kappa_0\\
   \frac{\kappa_a^2}{2 \kappa_0^2} \mathcal{E} & -2 \kappa_0 \leq
\mathcal{E} \leq 2 \kappa_0 \\
  \frac{\kappa_a^2}{2 \kappa_0^2} \left(
\mathcal{E}-\sqrt{\mathcal{E}^2-4 \kappa_0^2} \right) &
\mathcal{E}>2\kappa_0
\end{array}
\right. \nonumber
\end{eqnarray}
The condition for the non-Hermitian Hamiltonian to possess a
real-valued spectrum (i.e. to avoid complex-valued energies arising
from bound states outside the continuum) is derived in Appendix B.
Precisely, let $\xi_{1,2}$ be the two roots of the second-order
algebraic equation
\begin{equation}
\xi^2+\frac{E_a}{\kappa_0} \xi+1-(\kappa_a/\kappa_0)^2=0.
\end{equation}
\begin{figure}
\includegraphics[scale=0.43]{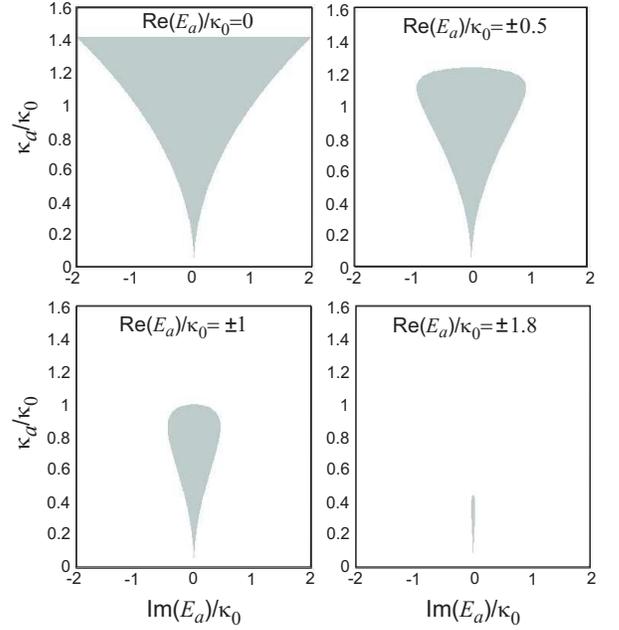} \caption{
Domains of non-existence of bound states for the Hamiltonian $H$ in
the $( \mathrm{Im}(E_a)/\kappa_0,\kappa_a/\kappa_0)$ plane (shaded
regions) for increasing values of the ratio
$|\mathrm{Re}(E_a)|/\kappa_0$. For a non-Hermitian Hamiltonian, i.e.
$\mathrm{Im}(E_a) \neq 0$, in the shaded regions the energy spectrum
of $H$ is real-valued and purely continuous. Spectral singularities
occur at the boundary of the shaded regions.}
\end{figure}
Then the Hamiltonian $H$ has a real-valued energy spectrum if and
only if $|\xi_{1,2}| \leq 1$. Figure 2 shows the domain in the plane
$( \mathrm{Im}(E_a)/\kappa_0,\kappa_a/\kappa_0)$ where $H$ has a
purely continuous energy spectrum for a few increasing values of the
ratio $|\mathrm{Re}(E_a)/\kappa_0|$. The domain lies in the sector
$\kappa_a/\kappa_0 \leq \sqrt 2 $ and shrinks toward
$\mathrm{Im}(E_a)/\kappa_0=\kappa_a/\kappa_0=0$ as $|\mathrm
{Re}(E_a)/\kappa_0| \rightarrow 2^-$.  For $|\mathrm{
Re}(E_a)/\kappa_0| \leq  2$,  bound states do exist for any value of
$\kappa_a/\kappa_0$ and $\mathrm {Im}(E_a)/\kappa_0$. The wider
domain is attained for $\mathrm{Re}(E_a)=0$. In particular, for
$\mathrm{Re}(E_a)=0$ and $\kappa_a/\kappa_0= \sqrt 2$, from Eq.(52)
it follows that $H$ has a real-valued energy spectrum provided that
\begin{equation}
-2 \kappa_0  < \mathrm{Im}(E_a) < 2 \kappa_0.
\end{equation}
Let us now consider the occurrence of spectral singularities.
According to Eqs.(19) and (20) and using Eqs.(49) and (51), a
spectral singularity at energy $\mathcal{E}=\mathcal{E}_0$, inside
the interval $(-2 \kappa_0, 2 \kappa_0)$, is found provided that the
following two equations are simultaneously satisfied
\begin{eqnarray}
\mathrm{Im}(E_a) = \pm \frac{\kappa_a^2}{\kappa_0} \sqrt{1-\left(
\frac{\mathcal{E}_0}{2 \kappa_0} \right)^2} \\
 \mathrm{Re}(E_a)= \left(1-\frac{\kappa_a^2}{2 \kappa_0^2} \right) \mathcal{E}_0.
\end{eqnarray}
For arbitrarily given values of $E_a$, $\kappa_a$ and $\kappa_0$,
the above conditions are generally not satisfied [nowhere for
$\mathcal{E}_0$ in the range $(-2 \kappa_0,2\kappa_0)$], i.e. the
non-Hermitian FFA Hamiltonian is generally diagonalizable. Spectral
singularities appear solely when a constraint among $\mathrm
{Re}(E_a)/\kappa_0 $, $\mathrm {Im}(E_a)/\kappa_0$ and
$\kappa_a/\kappa_0$ is satisfied. Let us first assume $\kappa_a/
\kappa_0$ strictly smaller that $\sqrt{2}$. In this case, a single
spectral singularity, at the energy
$\mathcal{E}_0=\mathrm{Re}(E_a)/(1-\kappa_a^2/2 \kappa_0^2)$ [see
Eq.(55)], is found provided that
\begin{equation}
\mathrm{Im}^2(E_a)=\frac{\kappa_a^4}{\kappa_0^2} \left[
1-\frac{\mathrm{Re}^2(E_a)}{(2
\kappa_0-\kappa_a^2/\kappa_0)^2}\right].
\end{equation}
It can be readily shown that the condition (56) defines the boundary
of the domains shown in Fig.2, i.e. a spectral singularity appears
when the boundary of existence of bound states is approached. The
case $\kappa_a/ \kappa_0 = \sqrt{2}$ is somehow singular. From
Eqs.(54) and (55), for $\kappa_a/ \kappa_0 = \sqrt{2}$ it follows
that there are {\it two} spectral singularities at energies
\begin{equation}
\mathcal{E}_0= \pm \sqrt{4 \kappa_0^2-\mathrm{Im}^2(E_a)}
\end{equation}
provided that $\mathrm{Re}(E_a)=0$.
 The physical meaning of such
spectral singularities will be discussed in Sec.III.C.

\subsection{Lattice realization}
FFA models are often encountered in connection to single-particle
electronic or photonic transport in one-dimensional tight-binding
lattices or networks (see, e.g.,
\cite{Mahan,Cini,Gadzuk00,Stefanou,Tanaka06,Fan98,Xu00,Longhi06,Longhi07,LonghiZeno,Kivsharun}
and references therein), and in most cases the underlying
Hamiltonian is Hermitian. In particular, the Hermitian limit of the
FFA Hamiltonian $H$ considered in the previous subsection has been
previously studied in \cite{Tanaka06,Longhi07} and shown to be
equivalent to a tight-binding Hamiltonian of a semi-infinity lattice
with an impurity site. The equivalence can be proven after
representing the Bloch states $|k\rangle$ of the tight-binding
energy band in terms of localized Wannier states $|n \rangle$ on a
lattice. Let us introduce the Wannier states $|n \rangle$ as
\begin{equation}
|n \rangle=\sqrt{\frac{2}{\pi}} \int_{0}^{\pi}dk \sin(nk) |k \rangle
\end{equation}
for $n=1,2,3,...$. Taking into account that
\begin{equation}
\int_{0}^{\pi} dk \; \sin(nk) \sin(m k)= \frac{\pi}{2} \delta_{n,m}
\end{equation}
\begin{figure}
\includegraphics[scale=0.7]{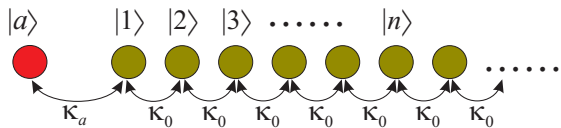} \caption{
(color online) Schematic of a semi-infinite one-dimensional
tight-binding lattice attached to a boundary impurity site $|a
\rangle$.}
\end{figure}
($n,m \geq 1$), one can readily show that the Wannier states form an
orthonormal system, i.e. $ \langle n | m \rangle =\delta_{n,m}$.
Additionally, from Eq.(58) it follows that the Bloch states $|k
\rangle$ can be decomposed as a superposition of Wannier states $|n
\rangle$ according to
\begin{equation}
|k \rangle = \sqrt{\frac{2}{\pi}} \sum_{n=1}^{\infty} \sin(nk) |n
\rangle.
\end{equation}
In the Wannier representation, one can readily show that
\begin{equation}
\int dk E(k) |k \rangle \langle k|=-\kappa_0 \sum_{n=1}^{\infty}
\left( |n \rangle \langle n+1|+ |n+1 \rangle \langle n| \right)
\end{equation}
and
\begin{equation}
V=\int dk  \left[ v(k) |a \rangle\langle k |+v^*(k) |k
\rangle\langle a | \right]=-\kappa_a(|a \rangle \langle 1|+|1
\rangle \langle a|)
\end{equation}
so that the Hamiltonian $H=H_0+V$ can be written in the equivalent
form
\begin{eqnarray}
H & = & - \kappa_0 \sum_{n=1}^{\infty} (|n \rangle \langle n+1|+|n+1
\rangle  \langle n |) + E_a |a \rangle
\langle a|+ \nonumber \\
& & - \kappa_a (|a \rangle  \langle  1 |+|1 \rangle \langle a |).
\end{eqnarray}
In its present form, Eq.(63) describes single-particle electron or
photon transport on a one-dimensional semi-infinite tight-binding
lattice \cite{Tanaka06,Longhi07,LonghiZeno}, with a hopping rate
$\kappa_0$ between adjacent sites of the lattice and with the
boundary attached to an impurity site $|a \rangle$ with 'complex'
potential energy $E_a$ and with hopping rate $\kappa_a$ (see Fig.3).
From a physical viewpoint, the complex potential at the boundary
impurity site may account for e.g. loss of the quantum particle flux
into other decay channels (quantum absorbing potentials
\cite{Marsiglio}), or optical gain or loss of light waves in
photonic structures \cite{Xu00,Longhi06}. For instance, light
transport in a semi-infinite waveguide array as in
Ref.\cite{LonghiZeno}, but with a lossy (or active) boundary
waveguide, provides a simple and experimentally accessible
realization of the tight-binding model (63). It should be noted that
transport and scattering phenomena in tight-binding lattices with
complex potentials have been theoretically investigated in recent
works (see, for instance, \cite{Marsiglio,Shapiro09}), however
spectral singularities were not found in these previous models.

\subsection{Spectral singularities and lattice wave reflection}

The physical meaning of spectral singularities in the non-Hermitian
FFA model can be captured by analyzing the wave reflection
properties of the lattice of Fig.3. As it will be shown below, a
spectral singularity corresponds to the appearance of a diverging
peak in the reflectance spectrum when the boundary site is an
'amplifying' impurity, i.e. when $\mathrm{Im}(E_a)>0$, and to the
vanishing of wave reflection when the boundary site is an
'absorbing' impurity, i.e. when $\mathrm{Im}(E_a)<0$. In the former
case, we retrieve for a 'discrete' scattering problem the physical
explanation of spectral singularities as resonances with vanishing
spectral width, established by Mostafazadeh
Ref.\cite{Mostafazadeh09JPA} in the framework of 'continuous' wave
scattering by complex potentials. Conversely, the latter case, i.e.
that of an 'absorbing' impurity site, gives a different
manifestation of a spectral singularity: a wave incident on the
lattice boundary is totally absorbed.\\
To analyze the reflection properties of the lattice shown in Fig.3,
let us expand the state vector $|\psi(t)\rangle$ as
\begin{equation}
|\psi(t)\rangle=c_a(t)|a \rangle +\sum_{n=1}^{\infty}c_n(t) |n
\rangle
\end{equation}
\begin{figure}
\includegraphics[scale=0.44]{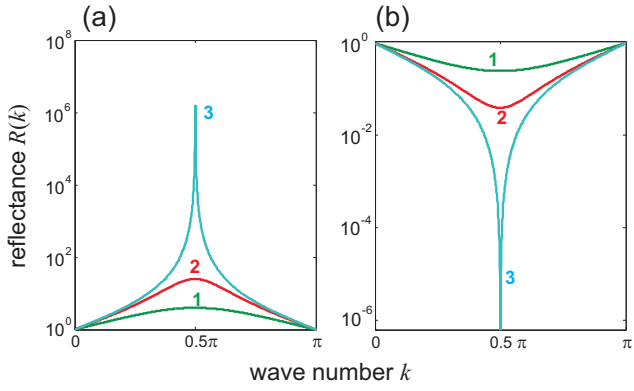} \caption{
(color online) Probability of particle reflection $R(k)$ for (a) an
amplifying impurity site [$\mathrm{Im}(E_a)>0$], and (b) an
absorbing impurity site [$\mathrm{Im}(E_a)<0$] for
$\kappa_a/\kappa_0=1$ and $\mathrm{Re}(E_a)=0$. Curves 1, 2 and 3
correspond to $|\mathrm{Im}(E_a)|/\kappa_a=1/3$, 2/3 and 1,
respectively. According to Eqs.(54) and (55), a spectral singularity
occurs at $k=\pi/2$ (i.e., $\mathcal{E}_0=-2 \kappa_0 \cos k=0$) for
$|\mathrm{Im}(E_a)|/\kappa_a=1$ (curve 3)}.
\end{figure}
\begin{figure}
\includegraphics[scale=0.44]{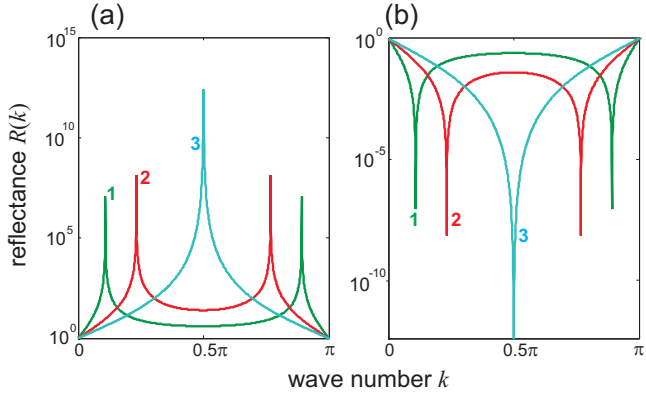} \caption{
(color online) Same as Fig.4, but for for $\kappa_a/\kappa_0=\sqrt
2$ and $\mathrm{Re}(E_a)=0$. Curves 1, 2 and 3 correspond to
$|\mathrm{Im}(E_a)|/\kappa_a=2/3$, 4/3 and 2, respectively.
According to Eq.(57), there are two spectral singularities,
corresponding to the energies  $\mathcal{E}_0=\pm [4
\kappa_0^2-\mathrm{Im}^2(E_a)]^{1/2}$, which coalesce at
$|\mathrm{Im}(E_a)|/\kappa_a=2$ (curve 3).}
\end{figure}
where $|c_n(t)|^2$ is the occupation probability of site $|n
\rangle$ and  $|c_a(t)|^2$ the occupation probability of the
boundary impurity site $|a \rangle$. From Eqs.(3) and (63), it
follows that
 the occupation amplitudes $c_n$ and $c_a$ satisfy the following
 coupled equations
 \begin{eqnarray}
 i\frac{dc_n}{dt} & = & -\kappa_0 (c_{n+1}+c_{n-1}) \; \; n \geq 2
 \\
 i\frac{dc_1}{dt} & = & -\kappa_0 c_2-\kappa_a c_a \\
 i\frac{dc_a}{dt} & = & -\kappa_a c_{1}+E_ac_{a}.
 \end{eqnarray}
Plane wave solutions to Eqs.(65-67) with wave number (momentum) $k$
($0 \leq k \leq \pi$), corresponding to eigenstates of $H$ with
energy $E(k)=-2 \kappa_0 \cos k$, are of the form
 $c_n(t)=\bar{c}_n(k) \exp(-iE(k)t)$,
$c_a(t)=\bar{c}_a(k) \exp(-iE(k)t)$, where
\begin{eqnarray}
\bar{c}_n(k) & = & \exp[-ik(n-1)]+r(k) \exp[ik(n-1)] \\
\bar{c}_a(k) & = & \frac{\kappa_0}{\kappa_a} [\exp(ik)+r \exp(-ik)]
\end{eqnarray}
and
\begin{equation}
r(k)=-\frac{\kappa_a^2-\kappa_0^2(2 \cos k +E_a /
\kappa_0)\exp(ik)}{\kappa_a^2-\kappa_0^2(2 \cos k +E_a /
\kappa_0)\exp(-ik)}
\end{equation}
is the spectral reflection coefficient (see, for instance,
\cite{LonghiZeno}). Note that $c_n(t)$ is given by the superposition
of two traveling waves, a regressive wave $ \exp[-ikn-iE(k)t]$ that
propagates along decreasing values of $n$, and a progressive wave $
\exp[ikn-iE(k)t]$ that propagates in the opposite direction, i.e.
which is reflected from the lattice boundary \cite{note5}. The
probability of reflection from the lattice boundary is given by
\begin{widetext}
\begin{equation}
R(k)=|r(k)|^2=\frac{[(\kappa_a^2-2\kappa_0^2) \cos k -\kappa_0
\mathrm{Re}(E_a)]^2+[\kappa_a^2 \sin k + \kappa_0
\mathrm{Im}(E_a)]^2}{[(\kappa_a^2-2\kappa_0^2) \cos k -\kappa_0
\mathrm{Re}(E_a)]^2+[\kappa_a^2 \sin k - \kappa_0
\mathrm{Im}(E_a)]^2}
\end{equation}
\end{widetext}
In the Hermitian limit ($\mathrm{Im}(E_a)=0$), one has $R(k)=1$,
i.e. the incident wave is completely reflected from the lattice
boundary. This is merely a consequence of conservation of the
particle probability in the scattering process. Conversely, for a
complex-valued energy $E_a$ of the impurity site, one has $R(k) \leq
1$ for $\mathrm{Im}(E_a) <0 $, i.e. for an absorbing potential, and
$R(k) \geq 1$ for $\mathrm{Im}(E_a) > 0$, i.e. for an amplifying
potential. In particular, for $\mathrm{Im}(E_a) > 0$ the reflection
probability $R(k)$ goes to infinity at wave numbers $k=k_0$ (with $0
\leq k_0 \leq \pi$) that satisfy simultaneously the two conditions
\begin{eqnarray}
\kappa_a^2 \sin k_0 & = & \kappa_0 \mathrm{Im}(E_a) \\
(\kappa_a^2-2 \kappa_0^2) \cos k_0 & = & \kappa_o \mathrm{Re}(E_a).
\end{eqnarray}
Physically, the condition $R \rightarrow \infty$ implies the
existence of an outgoing wave in the lattice that is sustained by
the amplifying complex potential at the impurity site. Such a
divergence of $R(k)$ is the signature of a spectral singularity of
$H$. In fact, taking into account that the energy of
incident/reflected waves is $\mathcal{E}_0=-2 \kappa_0 \cos k_0$, it
can be easily shown that Eq.(73) is equivalent to Eq.(55), whereas
Eq.(72) is equivalent to Eq.(54) with the upper (positive) sign on
the right hand side. Therefore, for an amplifying impurity site
[$\mathrm{Im}(E_a)>0$], the condition $R \rightarrow \infty$ is
equivalent to the appearance of a spectral singularity. This
equivalence extends, to our scattering problem on a truncated
lattice, the general result shown by Mostafazadeh
\cite{Mostafazadeh09JPA}, suggesting to interpret spectral
singularities of a non-Hermitian Hamiltonian as resonances with
vanishing width. However, for an absorbing potential energy at the
impurity site, i.e. for $\mathrm{Im}(E_a)<0$, our lattice model
indicates that the appearance of spectral singularities has a
different physical interpretation. In fact, for $\mathrm{Im}(E_a)<0$
the reflection probability $R(k)$ is bounded from above and smaller
than one, which prevents $R(k)$ to diverge. However, in this case
$R(k)$ can vanish at wave numbers $k=k_0$ (with $0 \leq k_0 \leq
\pi$) that satisfy simultaneously the two conditions [see Eq.(71)]
\begin{eqnarray}
\kappa_a^2 \sin k_0 & = & - \kappa_0 \mathrm{Im}(E_a) \\
(\kappa_a^2-2 \kappa_0^2) \cos k_0 & = & \kappa_o \mathrm{Re}(E_a).
\end{eqnarray}
Note that Eq.(75) is equivalent to Eq.(55), whereas Eq.(74) is
equivalent to Eq.(54) with the lower (negative) sign on the right
hand side. Therefore, for an absorbing impurity site
[$\mathrm{Im}(E_a)<0$], the appearance of a spectral singularity is
equivalent to the vanishing of the reflection probability $R$: an
ingoing plane wave with wave number $k_0$ is fully absorbed by the
impurity site at the lattice edge. An an example, Fig.4 shows the
behavior of $R(k)$ for an amplifying [Fig.4(a)] and for an absorbing
[Fig.4(b)] impurity site for $\mathrm{Re}(E_a)=0$ and
$\kappa_a/\kappa_0=1$. The different curves in the figures refer to
different values of $\mathrm{Im}(E_{a})/ \kappa_0$. Note that, at
the value of $\mathrm{Im}(E_{a})/ \kappa_0$ corresponding to the
appearance of the spectral singularity, a divergence and a zero in
the $R(k)$ curve are observed in Figs.4(a) and 4(b), respectively.
Figure 5 shows the behavior of $R(k)$ as in Fig.4, but for
$\mathrm{Re}(E_a)=0$ and $\kappa_a/\kappa_0= \sqrt 2$. In this case
there are two spectral singularities at energies given by Eq.(57),
which explain the two peaks [Fig.5(a)] or dips [Fig.5(b)] in the
reflectance curve $R(k)$. By increasing $|\mathrm{Im}(E_{a})|/
\kappa_0$, the two peaks (or dips) get closer each other, until they
coalesce at $|\mathrm{Im}(E_{a})|/
\kappa_0=2$ (curve 3 \cite{notecoale}).\\
\begin{figure}
\includegraphics[scale=0.43]{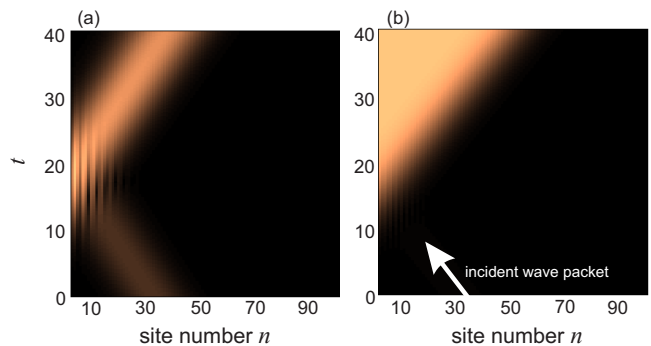} \caption{
(color online) Wave packet reflection in a semi-infinite lattice
with an amplifying impurity site for $\kappa_0=\kappa_a=1$,
$\mathrm{Re}(E_a)=0$ and $\mathrm{Im}(E_a)=1$. The initial wave
packet is Gaussian- shaped, with peak amplitude $|c_n(0)|=1$ at
$n=n_0=30$, width $\Delta n=12$ and momentum $k=3 \pi /10$ in (a),
and $k=\pi/2$ in (b) (corresponding to the spectral singularity
energy $\mathcal{E}_0=0$). In (b)the incident wave packet, indicated
by an arrow, is not visible owing to the large amplification of the
reflected wave.}
\end{figure}
\begin{figure}
\includegraphics[scale=0.43]{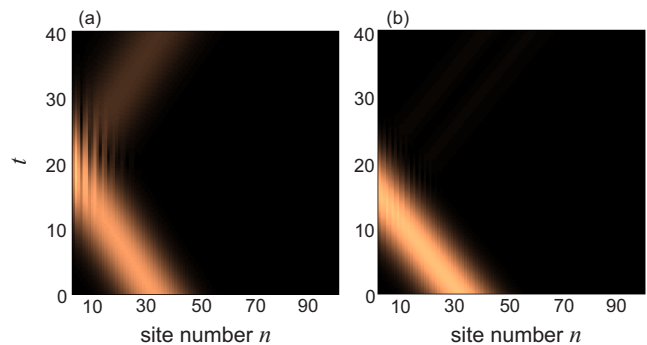} \caption{
(color online) Same as Fig.6, but for an absorbing impurity site
$\mathrm{Im}(E_a)=-1$ (other parameters as in Fig.6).}
\end{figure}
 Figures 6 and 7 show two
examples of wave packet reflection from the lattice boundary for an
amplifying (Fig.6) and an absorbing (Fig.7) impurity site. The
figures show a snapshot of $|c_n(t)|$ as obtained by numerical
analysis of Eqs.(65-67) assuming at $t=0$ a broad Gaussian
distribution of site occupation amplitudes, i.e.
$c_n(0)=\exp[-(n-n_0)^2/\Delta n^2-ik n]$, where $\Delta n$ is the
wave packet width, $k$ the mean wave packet momentum, and $n_0 \gg
\Delta n$ the wave packet center of mass. The large (diverging)
amplification of the reflected wave packet in Fig.6(b), and the
almost absence of wave packet reflection in Fig.7(b), are clearly
visible when the energy $\mathcal{E}=-2\kappa_0 \cos k$ of the
incoming wave packet attains the spectral singularity point
$\mathcal{E}_0=0$.\\
\begin{figure}
\includegraphics[scale=0.41]{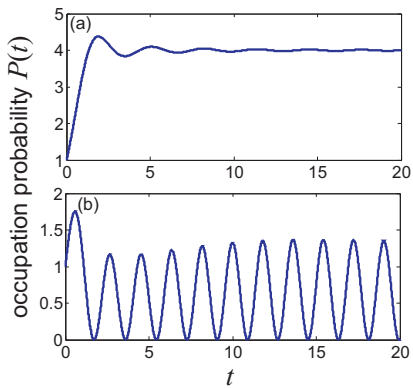} \caption{
(color online) Evolution of occupation probability $P(t)=|c_a(t)|^2$
in a semi-infinite lattice with an amplifying impurity site for
$\mathrm{Re}(E_a)=0$, $\mathrm{Im}(E_a)=1$ and for (a)
$\kappa_0=\kappa_a=1$, (b) $\kappa_0=1$, $ \kappa_a=\sqrt 2$.}.
\end{figure}

\begin{figure}
\includegraphics[scale=0.41]{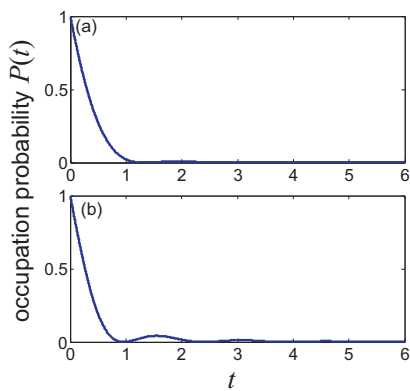} \caption{
(color online) Same as Fig.8, but for an absorbing impurity site
($\mathrm{Im}(E_a)=-1$; other parameters as in Fig.8)}.
\end{figure}
The interplay between spectral singularities and decay dynamics,
discussed in Sec.II.D, is exemplified in Figs.8 and 9. The system of
Eqs.(65-67) has been numerically integrated with the initial
condition $c_a(0)=1$ and $c_n(0)=1$. Note that, in the photonic
realization of the semi-infinite lattice model of
Ref.\cite{LonghiZeno}, such an initial condition simply corresponds
to initial excitation of the boundary waveguide. The behavior of the
site occupation probability $P(t)=|c_a(t)|^2$ for an amplifying and
for an absorbing impurity site is depicted in Figs.8 and 9,
respectively. Note that, according to the general analysis presented
in Sec.II.D, the survival probability $P(t)=|c_a(t)|^2$ decays to
zero in the absorbing case (Fig.9), and the existence of a spectral
singularity does not basically influence the decay dynamics of state
$|a \rangle$. Conversely, for an amplifying impurity site the
occupation probability $P(t)$ does not decay to zero. Note that for
$\kappa_a/\kappa_0 < \sqrt 2$ [like in Fig.8(a)], $P(t)$ converges
to a steady state value, whereas for $\kappa_a/\kappa_0 =\sqrt 2$
the probability $P(t)$ is an oscillating function [see Fig.8(b)].
The different behavior in the two cases is explained by observing
that, for $\kappa_a/\kappa_0 \neq \sqrt 2$ there is one spectral
singularity and thus in Eq.(45) there is only one pole that
contributes to the non-decaying part of $c_a(t)$. On the other hand,
for the somehow special case $\kappa_a/\kappa_0 = \sqrt 2$ the
Hamiltonian $H$ has two spectral singularities [see Eq.(57)], and in
Eq.(45) there are two poles that contribute to the non-decaying part
of $c_a(t)$. The interference of these two non-decaying terms
explains the oscillatory behavior of $P(t)$ in Fig.8(b) (see also
\cite{Longhi06}).

\section{Conclusions}
In this work a comprehensive analysis of the spectral properties of
a non-Hermitian extension of the Friedrichs-Fano-Anderson model has
been presented. The FFA model generally describes the decay of a
discrete state $|a \rangle$ of energy $E_a$ coupled to a continuum
of states. Here we have extended the ordinary model by allowing the
energy $E_a$ to become complex-valued, with either
$\mathrm{Im}(E_a)>0$ (the 'amplifying' case) or $\mathrm{Im}(E_a)<0$
(the 'absorbing' case). Contrary to the Hermitian FFA, it has been
shown by a direct diagonalization procedure and by the analysis of
the resolvent operator that spectral singularities in the continuous
spectrum may exist for both the amplifying and the absorbing
non-Hermitian FFA model. The physical implications and relevance of
spectral singularities have been discussed, in particular with
reference to a tight-binding realization of the non-Hermitian FFA
model that describes single-particle electronic or photonic
transport in a semi-infinite lattice attached to an impurity site
with a complex energy. Different behaviors have been found for an
amplifying and for an absorbing impurity site, reflecting the
circumstance that the divergence of the resolvent $G(z)$ appears
when the spectral singularity is approached either from above or
from below of the complex energy plane. For an amplifying impurity
site, the spectral singularity manifests itself as a divergence of
the reflection probability from the lattice boundary, a result which
is the 'discrete' analogous of the general result recently
established by Mostafazadeh for wave scattering by complex
potentials in the framework of the continuous Schr\"{o}dinger
equation \cite{Mostafazadeh09PRL}. As compared to
Ref.\cite{Mostafazadeh09PRL}, here we have also clarified the
physical relevance of spectral singularities in the temporal domain,
showing that in the amplifying non-Hermitian FFA Hamiltonian the
spectral singularity of the resolvent prevents the decay of state
$|a \rangle$ into the continuum, in spite of the absence of bound
states. For an absorbing impurity site, we have shown that a
spectral singularity corresponds to a zero of the reflection
probability from the lattice boundary. This result is clearly not
observable in the problem of wave scattering from complex barriers
addressed in Refs. \cite{Mostafazadeh06JPA,Mostafazadeh09PRL}, where
the double-degeneracy of energy levels plays a major role and a
spectral singularity can never correspond to the total absorption of
the incident wave.\\
Owing to the importance of the FFA model in different areas of
physics, it is envisaged that the present analysis may stimulate
further theoretical and experimental studies aimed to investigate
the unique features of non-Hermitian physical systems. In
particular, recent results obtained in photonic systems based on
coupled waveguides or arrays of coupled waveguides with controlled
regions of optical gain and/or loss \cite{Christodoulides}, indicate
that engineered photonic systems might provide an accessible
laboratory tool to experimentally observe spectral singularities.

\appendix

\section{Matrix elements of the resolvent} In this Appendix we
derive the expressions of the matrix elements
$\mathcal{G}_{a,a}(z)=\langle a|G(z) a \rangle$,
$\mathcal{G}_{k,a}(z)=\langle k|G(z) a \rangle$,
$\mathcal{G}_{a,k}(z)=\langle a|G(z) k \rangle$ and
$\mathcal{G}_{k,k'}(z)=\langle k|G(z) k' \rangle$ of the resolvent
[Eqs.(6-9) given in the text].\\
From the identity $G(z)(z-H_0-V)=(z-H_0-V)G(z)=\mathcal{I}$ it
follows that
\begin{eqnarray}
 \langle a |G(z)(z-H_0) a\rangle- \langle a |G(z) V a\rangle & = & 1 \\
\langle a |G(z)(z-H_0) k \rangle- \langle a |G(z) V  k \rangle & = & 0 \\
\langle k |(z-H_0)G(z) a \rangle- \langle k |V G(z)   a \rangle & = & 0 \\
\langle k |G(z)(z-H_0) k' \rangle- \langle k | G(z) V   k' \rangle &
= & \delta(k-k').
\end{eqnarray}
Taking into account that
\begin{equation}
V|a\rangle=\int dk v^*(k)|k\rangle \; , \; V|k\rangle=v(k)|a\rangle
\end{equation}
and that $(z-H_0)|a\rangle=(z-E_a)|a\rangle$,
$(z-H_0)|k\rangle=(z-E(k))|k\rangle$, Eqs.(A1) and (A2) take the
form
\begin{eqnarray}
(z-E_a)\mathcal{G}_{a,a}(z)-\int dk v^*(k)\mathcal{G}_{a,k}=1 \\
(z-E(k))\mathcal{G}_{a,k}(z)-v(k)\mathcal{G}_{a,a}(z)=0
\end{eqnarray}
which can be solved for $\mathcal{G}_{a,a}$ and $\mathcal{G}_{a,k}$,
yielding Eqs.(6) and (7) given in the text. To calculate
$\mathcal{G}_{k,a}(z)$, we use Eq.(A3) and note that $\langle
k|(z-H_0)G(z) a\rangle=\langle (z^*-H_0^{\dag}) k|G(z)
a\rangle=(z-E(k))\mathcal{G}_{k,a}(z)$ and $\langle
k|VG(z)a\rangle=\langle V k |G(z) a \rangle =
v^*(k)\mathcal{G}_{a,a}(z)$. This yields
$(z-E(k))\mathcal{G}_{k,a}(z)-v^*(k)\mathcal{G}_{a,a}(z)=0$, which
can be solved for $\mathcal{G}_{k,a}(z)$, yielding Eq.(8) given in
the text. Finally, the matrix element $\mathcal{G}_{k,k'}(z)$ is
obtained from Eq.(A4), which can be written in the form $(z-E(k'))
\mathcal{G}_{k,k'}(z)- v(k') \mathcal{G}_{k,a}(z) = \delta(k-k')$,
i.e.
\begin{equation}
\mathcal{G}_{k,k'}(z)=\frac{v(k')
\mathcal{G}_{k,a}(z)}{z-E(k')}+\frac{\delta(k-k')}{z-E(k')}.
\end{equation}
Substitution of Eq.(8) into Eq.(A8) finally yields Eq.(9) given in
the text.

\section{Conditions for a real-valued energy spectrum of the non-Hermitian Hamiltonian}
In this Appendix we derive the necessary and sufficient conditions
that ensure a real-valued energy spectrum for the non-Hermitian FFA
Hamiltonian $H$ introduced in Sec.III.A. As shown in Sec.II.B, this
condition is equivalent to the vanishing of the point-spectrum of
$H$, i.e. to the absence of bound states. The detailed calculations
can be performed following two different, though equivalent,
approaches. The first one starts from the representation of $H$ in
the $\{ |a \rangle , |k \rangle \}$ basis (the Bloch basis), whereas
the second approach uses a different decomposition of $H$, namely on
the $\{ |a \rangle , |n \rangle \}$ basis, where $|n \rangle$ are
the Wannier states introduced in Sec.III.B (the Wannier basis). For
the sake of completeness, we
present the detailed calculations for both approaches.\\
\\
{\it 1. Bloch-basis representation of H}.
 As shown in Sec.II.B, the absence of bound states of $H$ requires that Eq.(18) does not
admit of any solution in the complex $z$ plane. Using the expression
(50) of the self-energy $\Sigma(z)$, Eq.(18) takes the form
\begin{equation}
\left( 1-\frac{\kappa_a^2}{2
\kappa_0^2}\right)z-E_a=-i\frac{\kappa_a^2}{2 \kappa_0^2}
\sqrt{4\kappa_0^2-z^2}.
\end{equation}
We can solve Eq.(B1) by introducing, in place of $z$, the new
complex-valued variable $\mu$ defined by
\begin{equation}
z=-\kappa_0[ \exp( \mu)+ \exp(-\mu)]=-2 \kappa_0 \cosh \mu.
\end{equation}
Without loss of generality, we may assume $\mathrm{Re}(\mu) > 0$. In
fact, the function $ z(\mu)$ defined by Eq.B(2) is invariant for the
inversion $\mu \rightarrow - \mu$, so that we may restrict our
analysis to the case $\mathrm{Re}(\mu) > 0$. With such a
substitution, the square root on the right hand side of Eq.(B1) can
be solved analytically, yielding $\pm 2 i \kappa_0 \sinh \mu$. Some
care should be taken when choosing the right determination (i.e.
sign) of the square root \cite{note4}. For $\mathrm{Re}(\mu) > 0$,
one obtains
\begin{equation}
2 \left( 1-\frac{\kappa_a^2}{2 \kappa_0^2}\right) \cosh
\mu+\frac{E_a}{\kappa_0}=-\frac{\kappa_a^2}{\kappa_0^2} \sinh \mu.
\end{equation}
After setting $\xi=\exp(\mu)$, from Eq.(B3) one obtains Eq.(52)
given in the text once $\cosh \mu$ and $\sinh \mu$ are expressed in
terms of the exponentials $\exp(\pm \mu)=\xi^{\pm 1}$. Therefore, if
the two roots $\xi_{1,2}$ of Eq.(52) satisfy the condition
$|\xi_{1,2}| \leq 1$, Eq.(B3) does not have roots with
$\mathrm{Re}(\mu) > 0$, and hence $H$ does not have bound
states.\\
\\
{\it 2. Wannier-basis representation of H}. In this approach, we use
the tight-binding representation of the Hamiltonian $H$ using the
Wannier function basis [Eq.(63)]. Bound states of $H$ correspond in
this case to surface states localized near the edge of the truncated
lattice of Fig.3. They can be directly determined by looking for a
solution to Eqs.(65-67) of the form
\begin{equation}
c_n(t)=\exp[-\mu(n-1)-iEt] , \; c_a(t)=A \exp(-iEt)
\end{equation}
($n \geq 1$), where $E$ is the energy of the surface state. The
constants $\mu$ and $A$, as well as the dependence of $E$ on $\mu$,
are readily determined by substituting Eq.(B4) into Eqs.(65-67). One
obtains
\begin{eqnarray}
E & = & -2 \kappa_0 \cosh \mu   \\
E & = & -\kappa_0 \exp(-\mu)-\kappa_a A \\
E A & = & -\kappa_a+E_a A.
\end{eqnarray}
from which the following second-order algebraic equations for $\xi=\exp(\mu)$ is readily obtained
\begin{equation}
\xi^2+\frac{E_a}{\kappa_0} \xi+1-\frac{\kappa_a^2}{\kappa_0^2}=0
\end{equation}
which is Eq.(52) given in the text. Localization of the surface
state at the lattice edge requires $c_n \rightarrow 0$ as $n
\rightarrow \infty$, i.e. $\mathrm{Re}(\mu)>0$ [see Eq.(B4)]. Therefore, if
the two roots $\xi_{1,2}$ of Eq.(B8) satisfy the condition $|\xi_{1,2}| \leq 1$, there are not surface states
at the lattice edge.


\begin{thebibliography}{99}
\bibitem{note1} For ${\rm Im}(E_a) = 0$, we recover the Hermitian
limit of the FFA Hamiltonian. In this case a divergence of the
resolvent for $z \rightarrow \mathcal{E}_0$, obtained from Eq.(20)
with $V(\mathcal{E}_0)=0$, corresponds to a bound state embedded
into the continuum (see, e.g., \cite{Sudarshan,Miyamoto05}) rather
than to a spectral singularity of $H$.
\bibitem{note2}
In fact, if $z$ were a real-valued root of Eq.(18) outside the
interval $(E_1,E_2)$, according to Eq.(12) one would have
$\Sigma(z)=\Delta(z)$ because $V(z)=0$. Hence, $\Sigma(z)$ turns out
to be real-valued. To satisfy Eq.(18), the condition
$\mathrm{Im}(E_a) = 0$ must be thus satisfied. This means that a
real-valued root $z$ of Eq.(18) necessarily requires $H$ to be
Hermitian.
\bibitem{note3} For a featureless continuum, for which
$\Delta(\mathcal{E})$ and $V(\mathcal{E})$ are smooth functions of
$\mathcal{E}$, the resonance curve (39) has a Lorentzian shape and
the quantum system, intially prepared in state $|a\rangle$, would
decay into the continuum following a (nearly) exponential decay law
with lifetime $1/ [\pi V(E_a)]$ (Weisskopf-Wigner or Breit-Wigner
approximation). See also the discussion in Sec.II.D.
\bibitem{noteq}
We assume here that the spectrum of $H$ is purely continuous, i.e.
we assume that there are not bound states (outside or embedded in
the continuum). In this case, for the Hermitian FFA model the decay
of the survival probability $P(t)$ is always complete.
\bibitem{note4}
The proper determination of the square root on the right hand side
of Eq.(50) must be chosen such that $\mathrm{Im} \left(
\Sigma(z=\mathcal{E} \pm i 0^+) \right)=\mp \pi V(\mathcal{E})$,
according to Eq.(12).
\bibitem{notecoale}
At the coalescent point, the resolvent $G(z)$ has a second-order
pole at $z=0$.
\end{thebibliography}
\end{document}